\documentclass[aps, prl, showpacs, amsmath, twocolumn, groupedaddress, superscriptaddress]{revtex4}
\usepackage{graphics}
\usepackage{graphicx}
\usepackage{subfigure}
\usepackage{float}
\usepackage{longtable}
\usepackage{amssymb}

\newcommand{\ca}{$\mathrm{Ca^{+}\,}$}

\newcommand{\mgh}{$\mathrm{MgH^{+}\,}$}

\newcommand{\Ket}[1]{\left\lvert{#1}\right\rangle}
\newcommand{\vib}[1]{\Ket{#1}_\text{Tr}}

\newcommand{\mol}[1]{\Ket{#1}_{\text{mol}}}

\newcommand{\gShort}[2]{$\Ket{g}_\text{at}\Ket{#1}_\text{mol}\Ket{#2}_{\text{Tr}}$}
\newcommand{\shelveShort}[2]{$\Ket{s}_\text{at}\Ket{#1}_\text{mol}\Ket{#2}_{\text{Tr}}$}
\newcommand{\unShelveShort}[2]{$\Ket{g}_\text{at}\Ket{#1}_\text{mol}\Ket{#2}_{\text{Tr}}$}
\newcommand{\cas}{\mathrm{\Ket{g}_\text{at}}}
\renewcommand{\cap}{\mathrm{\Ket{e}_\text{at}}}

\newcommand{\caShelve}{\mathrm{\Ket{s}_\text{at}}}

\begin{document}
\title{Probabilistic state preparation of a single molecular ion by projection measurement}
\author{I. S. Vogelius}
\affiliation{Department of Physics and Astronomy,
  University of  Aarhus, 8000 {\AA}rhus C, Denmark}
\author{L. B. Madsen}
\affiliation{Department of Physics and Astronomy,
  University of  Aarhus, 8000 {\AA}rhus C, Denmark}
\author{M. Drewsen}
\affiliation{Department of Physics and Astronomy,
  University of  Aarhus, 8000 {\AA}rhus C, Denmark}
\affiliation{QUANTOP - Danish National Research Foundation Center for
  Quantum Optics, Department of Physics and Astronomy, University of Aarhus,
  8000 {\AA}rhus C, Denmark}

\date{\today}
\begin{abstract}

We show how to prepare a single molecular ion in a specific internal quantum state in a
situation where the molecule is trapped and sympathetically cooled by an atomic ion and
where its internal degrees of freedom are initially in thermal equilibrium with the
surroundings. The scheme is based on conditional creation of correlation between the
internal state of the molecule and the translational state of the collective motion of
the two ions, followed by a projection measurement of this collective mode by atomic ion
shelving techniques. State preparation in a large number of internal states is possible.

\end{abstract}

\pacs{33.80.Ps,33.20.Vq,82.37.Vb}

\maketitle

Investigations of  production and trapping of cold neutral and ionic
molecules~\cite{MeijerTrappingOH,EurPhysJDreviewColdMolMasneou,egorov} point to a wealth
of possible applications, including studies of molecular Bose-Einstein
condensates~\cite{KetterleMolecularBEC_new,BEC_BCSCrossFermiAtoms,BEC_BCSTransInLi6ENSGroup,BEC_BCSBurnett,BEC_BCSJin},
investigations of collision and reaction dynamics at low temperature~\cite{Smith1994},
high-resolution spectroscopy~\cite{DoyleSpectroscopyCaFCooled}, coherent control
experiments~\cite{Rice2000}, and state specific reactions
studies~\cite{NatureQuantStateReaction,ScienceQuantStateResolveReaction}. For much of
this research, long-term localized and state-specific targets are highly desirable. One
way to obtain such targets is to work with trapped molecular ions sympathetically cooled
by atomic ions where previous investigations show that molecular ions can be
translationally cooled to temperatures of a few mK, at which stage they become immobile
and localize spatially in Coulomb crystal
structures~\cite{molhave,JensLykkeMassMeasurements}. Though these molecules are
translationally cold, studies indicate that the internal degrees of freedom of at least
smaller hetero-nuclear molecules, due to their interaction with the black-body radiation
(BBR), are close to be in equilibrium with the temperature of the
surroundings~\cite{BertelsenRotationalTempMeasurement}. This is not unexpected since the
many order of magnitude difference between the internal transitions frequencies in the
molecule ($\gtrsim 10^{11}\,$Hz) and the frequency of the collective vibrational modes in
the Coulomb crystals ($\lesssim 10^7\,$Hz) leads to very inefficient coupling between
these degrees of freedom. Several schemes were recently proposed to cool the rotational
temperature of translationally cold heteronuclear molecular
ions~\cite{vogelius,vogeliusLongPra}.

Here, we focus on an alternative route to the production of molecular ions in specific
states. The physical system used for this purpose consists of one trapped molecular ion
sympathetically cooled by a simultaneously trapped atomic ion. Such a situation was
previously realized and it was shown to be possible to determine the molecular ion
species non-destructively by a classical resonant excitation of one of the two axial
collective modes of the two-ion system~\cite{JensLykkeMassMeasurements}. With this setup,
we now propose to exploit the quantum aspect of the same collective modes to create
correlations between the internal state of the molecular ion and the collective motional
state in the trap potential. Previously, correlations in two-ion systems were essential
in, e.g., demonstrations of quantum logical gates~\cite{BlattGroupCaTwoIonGate} and in a
proposal for high-resolution spectroscopy~\cite{Wineland2003}.

\begin{figure}[ttb]
  \includegraphics[width=0.45\textwidth]{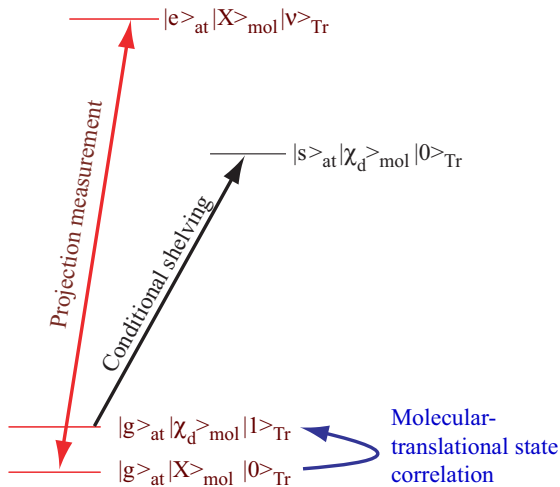}\\
\caption{(color online) Idealized sketch of the state-selection
sequence. The first step is cooling to the ground state of the
collective translational mode, $\vib{0}$ with the molecule in the
electronic ground state and with the internal rovibrational states
characterized by a statistical ensemble $\mol{X}$. The second step
is to correlate the collective mode of the two-ion system with the
internal state of the molecule by exciting the collective mode to
$\vib{1}$ if the molecule is in a specific internal state,
$\mol{\chi_\text{d}}$. Next, if the collective mode is excited,
conditional shelving transfers the atom to the long-lived
metastable state, $\caShelve$. Finally, the projection measurement
proceeds by exposing the atomic ion to light which is resonant
with the transition between the atomic ground state, $\cas$, and
an excited state, $\cap$. Rescattered light will then be absent
only if the atom is in the shelved state and hence the the
molecule in the desired internal state
$\mol{\chi_\text{d}}$.}\label{fig:Scheme}
\end{figure}

As depicted in Fig.~\ref{fig:Scheme}, the state preparation of the molecular ion ideally
involves the following steps. First, the two-ion system is cooled to its collective
motional ground
state~\cite{SidebandCoolingTwoIonCaFromBlattGroup,WinelandTwoSpeciesGroundStateCooling}
with the molecule in the electronic ground state and with a Maxwell-Boltzmann
distribution over rovibrational states. We consider only one of the two independent axial
modes of the two-ion system and refer to it as the collective mode. Second, laser fields
are applied to induce transitions between the ground and the first excited motional
states conditioned on the specific rovibrational state of the molecular ion. This
procedure creates correlations between the motional state of the two-ion system and the
internal state of the molecular ion. Next, conditioned on an excitation of the collective
mode, an atomic shelving transition to a metastable state is driven by another laser
field. Finally, laser fields are applied to project the atomic ion on the shelved
(non-fluorescing) or non-shelved (fluorescing) state. If no fluorescence is observed, we
conclude that the molecular ion is in the internal state of interest. Contrary, if
fluorescence is present, the ion is not in the desired state. In the latter case, after a
duration of time sufficiently long to bring the molecule back in thermal equilibrium
(typically through interaction with BBR), the procedure is repeated. Eventually, no
fluorescence is detected in the final step, and the molecular ion is known to be in the
desired quantum state. A state-to-state analysis of the procedure is presented
schematically in Table I.

\begin{table}[!tb]
\begin{center}
\begin{tabular}{|c|c|c|}
IE &\multicolumn{2}{c|}{\gShort{X}{0}}\\
\hline CEI  & \gShort{\chi_\text{d}}{1} & \gShort{X'}{0} \\
\hline CS & \shelveShort{\chi_\text{d}}{0} & \gShort{X'}{0} \\
\hline PM & dark &  fluorescent\\  \hline
\end{tabular}
\end{center}
\caption{ \label{TBL:SelectionSequence} The evolution of the system through the
state-selection sequence depicted in Fig.~\ref{fig:Scheme}. IE: Initialization of the
External state of the two-ion system by cooling the collective mode to the ground state.
CEI: Correlation of collective External modes and Internal molecular state. CS:
Conditional Shelving. PM: Projection Measurement. In the table, $\mol{X'}$ denotes the
modified statistical ensemble of internal molecular states when the desired state,
$\mol{\chi_d}$, has been removed from $\mol{X}$. The portion of the ensemble of
rovibrational states that was initially in the desired molecular state $|\chi_d
\rangle_{\text{mol}}$ is conditional shelved and unaffected by the projection
measurement. The remaining rovibrational states, on the other hand, are unaffected by the
conditional shelving and therefore lead to fluorescence upon projection measurement.}
\end{table}

Before a discussion of a realistic implementation of the proposed
scheme, we evaluate the effect of imperfections in the various
steps of the procedure.

1) \emph{Initialization of the external state of the two-ion system (IE)}. Cooling of a
two-ion system completely to the motional ground state, $| 0\rangle_\text{Tr}$, is
unrealistic, but several experiments show that it is feasible to achieve $W_0 = 95$$\%$
population in $| 0 \rangle_\text{Tr}$ in the case of two atomic ion
species~\cite{WinelandTwoSpeciesGroundStateCooling,SidebandCoolingTwoIonCaFromBlattGroup}.
The same degree of cooling is expected for an atomic-molecular ion system. We therefore
use a Boltzmann distribution with $W_0=95\%$, and this results in $W_1\simeq 4.7\%$ and
$W_2\simeq 0.3\%$ for the populations in $| 1 \rangle_\text{Tr}$ and $|2
\rangle_\text{Tr}$, respectively.


2) \emph{Correlation of external motion and the internal molecular state (CEI)}. This
part of the procedure can, e.g., be accomplished by inducing transitions between
rotational sub-states of the molecule using Raman
$\pi$-pulses~\cite{BlattGroupCaTwoIonGate,MonroeWinelandDemonstrationOfQuantLogicGateInTrap}
in a way similar to that demonstrated for sub-states in atomic
ions~\cite{MonroeWinelandDemonstrationOfQuantLogicGateInTrap}. Alternatively, if there
are no rotational sub-states, as in the case of J=0, or if for some reason it is desired
to stay in a specific sub-state, a sequence of two laser pulses can be applied like in
STIRAP processes\cite{STIRAPreviewArticle}. First, a pulse couples the final state,
$\mol{\chi_d}\vib{1}$, and intermediate states, $\mol{\chi'}\vib{\nu }$, with a coupling
strength characterized by the free molecule Rabi frequency, $\Omega _s (t)$, while a
delayed pulse couples the initial state, $\mol{\chi_d}\vib{0}$, to $\mol{\chi'}\vib{\nu
}$ with free molecule Rabi frequency $\Omega_p(t)$. Here, $\mol{\chi_d}$ and
$\mol{\chi'}$ denote the desired and intermediate molecular state, respectively. Though
the pulse sequence resembles a STIRAP process, there is an important difference since the
two laser pulses, that are only shifted in frequency by the collective mode frequency
($\simeq 10\,$MHz), do interact with same internal transitions of the molecule. To model
the effect of such a two-pulse process, we expand the state of the two-ion system as $|
\Psi(t) \rangle = \sum_{\nu_\text{Tr}=0}^{\nu_\text{Tr,max}} c_{\nu_\text{Tr}} (t)
\mol{\chi_d}\vib{\nu} + b_{\nu_\text{Tr}} (t)\mol{\chi'} \vib{\nu} $ with
$\nu_\text{Tr,max} = 5$ for convergence, initial condition $ c_0(t=0) = 1$ and desired
final state $\mol{\chi_d}\vib{1}$. In the simulations, both laser pulses are assumed to
be Gaussian in time with a width $\tau = 50$ $\mu$s(FWHM) and separated by $1.3\tau$.
Furthermore, we assume the intermediate molecular state to be a vibrational excited state
with a spontaneous decay rate of 100 s$^{-1}$. With a realistic molecular transition
wavelength of 4 $\mu$m, a maximum free molecule Rabi frequencies of 7 MHz for both
pulses, and a detuning from the intermediate state of $\delta \simeq 10\,$MHz, we find
that more than $80\,\%$ of the population can be transferred to motional excited states.
Of experimental importance, the transfer efficiency was found to be stable when varying
the detuning a few MHz~\footnote{Two laser pulses with a difference frequency which is
insensitive to fluctuations in the laser frequency may be generated from a single CW
laser source by application of acousto-optical modulators. The assumed Rabi frequency of
7 Mhz can be achieved by focusing CW laser beams with a modest power of $\sim 10$ mW to
spot sizes of $\sim 1$mm$^2$.}. In the following, we use the more conservative estimate
of the transfer efficiency, $\wp_\text{CEI} = 0.7$.

3) \emph{External state conditional shelving (CS)}. Shelving conditioned on excitation of
the collective mode can, e.g., be achieved by driving red side-band transitions between
motional states by $\pi$-pulses~\cite{MonroeWinelandDemonstrationOfQuantLogicGateInTrap}.
As an alternative, we consider the STIRAP type process from the atomic ground state,
$\Ket{g}_\text{at}$ to the shelved atomic state, $\Ket{s}_\text{at}$, via an intermediate
state, $\Ket{i}_\text{at}$. We expand the state of the combined system as $| \Psi(t)
\rangle = \sum_{\nu_\text{Tr}=0}^{\nu_\text{Tr,max}} \left( c_{\nu(t)_\text{Tr}} |
g\rangle_\text{at} \mol{\chi_d}\vib{\nu} + b_{\nu(t)_\text{Tr}} |
i\rangle_\text{at}\mol{\chi_d}\vib{\nu}\right.$\\$\left.+ a_{\nu(t)_\text{Tr}} |
s\rangle_\text{at}\mol{\chi_d}\vib{\nu} \right)$. As initial conditions we use the final
amplitudes from the CEI step, which showed significant population in the first few
excited motional states. Our simulations show a transfer efficiency $\wp_\text{CS}$ of
more than 95 \% is obtained even with population in several excited motional states. We
use the more conservative estimate $\wp_\text{CS} =0.7$ in the
following. 

 4) \emph{Projection measurement (PM)}.
The final step, projection measurement on the atomic ion, can be
made very efficient. With a typical exposure time $T=5\,$ms, one
should be able to determine the projected atomic state with more
than a 95 $\%$ confidence~\cite{Rowe01}.

-\emph{Over-all confidence of steps 1)-4).} The probability of being in the ideal initial
state $\Ket{g}_\text{at}\mol{\chi_d}\vib{0}$ is $P_{\chi_\text{d}}W_0 $, where
$P_{\chi_\text{d}}$ denotes the initial population in $\mol{\chi_d}$ and $W_\nu $ the
initial population in the collective motional state $\vib{\nu }$. Since this state has to
go through both CEI and CS to reach the shelved state, the probability of finding the
molecule in the shelved state after the selection sequence is $S_{prep}=P_{\chi_d} W_0
\wp_{CEI}\wp_{CS}$. Unfortunately, a false positive result can occur if the system is
initially in a state $\vib{\nu }$ with $\nu \geq 1$ since the system may proceed through
CS without exciting the motional state during CEI. The probability of a false positive
measurement is then $E=(1-P_{\chi_\text{d}})(1-W_{0})\wp_\text{CS}$, where the first two
factors account for the initial population in the excited states of the collective
motion, but not in $\mol{\chi_d}$, and the last factor accounts for the necessary
application of CS. We define the confidence of a measurement as $F= S_{prep} \big/
(S_{prep}+E)$.

As a test case we assume $P_{\chi_\text{d}}=5\%$ and use a thermal distribution over
external vibrational states determined by letting $W_0=95\, \%$. Non-fluorescence in the
final stage of the state preparation then give a confidence of the molecule being in
$\mol{\chi_d}$ of about $F\simeq 0.4$ which is too marginal for the procedure to be
useful.

5) \emph{State purification (SP)}.  The principal source of error is false positive
detections stemming from initial population in  $| 1 \rangle_\text{Tr}$. These errors are
excluded by use of the state-purification procedure presented in Table II. If no
fluorescence is detected after the state preparation process, CEI is reapplied. The
second step is another CS process transferring $| s \rangle_\text{at}$ to $|
g\rangle_\text{at}$ on the red sideband of the collective motion. Finally PM is repeated.
The desired final state is now flourescent while other states are dark.

\begin{table}[!tb]
\begin{center}
\begin{tabular}{|c|c|c|}
Start & \shelveShort{\chi_\text{d}}{0} &  \shelveShort{X'}{0} \\
\hline CEI & \shelveShort{\chi_\text{d}}{1}  & \shelveShort{X'}{0}
\\
\hline CS  & \unShelveShort{\chi_\text{d}}{0} & \shelveShort{X'}{0} \\
\hline PM & fluorescent & dark \\  \hline
\end{tabular}
\end{center}
\caption{\label{TBL:PostSelectionSequence} State Purification (SP). Start: The states of
the system ending in the shelved atomic state after the state preparation process 1)-4)
when assuming finite temperature of the collective mode. SP proceeds by a second
Correlation of the External mode and Internal molecular state (CEI), exciting the
collective mode if the molecule is in $\mol{\chi_\text{d}}$. Then Conditional Shelving
(CS) transfers the atom back to the ground state $\cas$ if the collective mode is
excited. Finally, a second Projection Measurement (PM) detects the atom in $\cas$ if the
molecule is in $\mol{\chi_\text{d}}$.}
\end{table}

The probability of a successful detection after the SP procedure is estimated by
$S_{pur}=S_{prep}\wp_{CEI}\wp_{CS}$, as the population has to go through both CEI and CS
after the state preparation. The probability of a false positive detection caused by
population initially in excited motional states is now given by $E1=W_{l\geq
2}(1-P_{\chi_\text{d}})\wp_{CS}^2$, since the system must start in the second excited
state or higher to pass CS twice without being excited during CEI. Another source of
false positive measurements appears due to stochastic heating during the relatively slow
PM process. Assuming a heating rate of $\Gamma=10$ vibrational quanta per second
\cite{SidebandCoolingTwoIonCaFromBlattGroup}, the error induced by stochastic heating is
$E2= T \Gamma(1-P_{\chi_\text{d}})W_1 \wp_{CS}^2$, since only population initially in
$\vib{1}$ will introduce error which is not accounted for in $E1$.

The confidence of state preparation after SP is defined as
$F'={S_{pur}}\big/({S_{pur}+E1+E2})$ and SP improves the confidence of state preparation
to more than $80\, \%$ (see Fig. 2). Note that if $W_0 =100\%$, the measurement cycle
leads to 100 \% certain state-preparation without SP, while increasing $\rho_{CEI}$ to
$100\%$ only leads to $10\%$ increase in the confidence of the final state preparation.
Finally, we note that the preparation efficiency is independent of the conditional
shelving efficiency, $\wp_{CS}$.

As an alternative to PS, we may perform CEI and CS on the second motional sideband, i.e.,
consider $\Ket{g}_\text{at}\mol{\chi_d}\vib{0} - \Ket{g}_\text{at}\mol{\chi_d}\vib{2}$ in
the CEI step and $\Ket{g}_\text{at}\mol{\chi_d}\vib{2} -
\Ket{s}_\text{at}\mol{\chi_d}\vib{0}$ in the CS step. Since the initial probability $W_2$
for being in $| 2\rangle_\text{Tr}$ is much smaller than $W_1$, the confidence of the
preparation is improved considerably without applying SP, and simply given by $F''=
S_{prep} \big/ (S_{prep}+E'')$ with $E'' = E\times (1-W_0-W_1)/(1-W_0)\sim E \times
0.06$. The required laser intensity would, however, increase significantly due to the
weaker coupling between $|0\rangle_\text{Tr}$ and $|2\rangle_\text{Tr}$ compared with the
$|0\rangle_\text{Tr}$ -- $|1\rangle_\text{Tr}$ coupling.

The average time needed for the performance of a successful detection depends partly on
the probability of finding the molecule in $\mol{\chi_d}$ at thermal equilibrium, partly
on the time needed for the projection measurement cycle including the purification step,
and partly on the time scale for re-thermalization. Even for polar diatomic molecules
with a relative large rotational constant, the time scale for re-thermalization
($\tau_\text{re} \sim$ 5 s) is much longer than the projection measurement cycle. Hence,
the average time is estimated by $\tau_\text{re} / S_{prep}$ which for a state with an
thermal population of $\sim$ 10 $\%$ is  $\sim 100$s.

\begin{figure}[ttb]
  \includegraphics[width=0.44\textwidth]{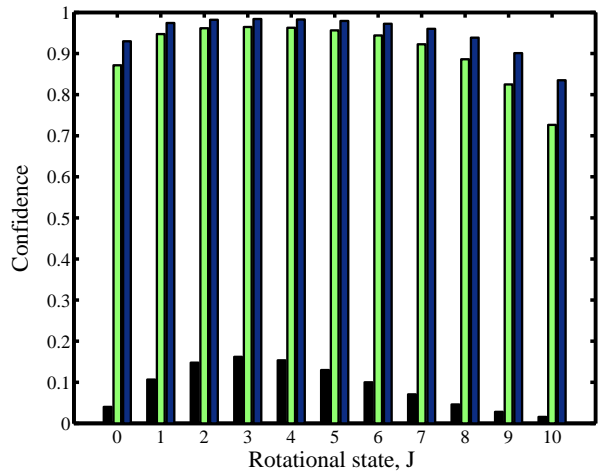}\\
\caption{(color online) Confidence of finding a \mgh ion in rotational states. The black
columns correspond to a $300\,$K Boltzmann distribution over rotational states. The light
grey (green) columns present the achievable confidence, $F'$, when state purification is
applied. Finally, the dark grey (blue) columns show the confidence, $F''$, obtainable
when selections using the second sideband are applied without
state-purification.}\label{fig:Selectivity}
\end{figure}

As a specific test case for an implementation of the state preparation scheme, we now
focus on a \mgh ion internally in equilibrium with BBR at 300 K and trapped together with
an atomic ion amenable for CS, e.g., a \ca ion. We neglect the vibrational quantum number
and take into account only the rotational states of the \mgh ion since at room
temperature it is in the vibrational ground state with more than 99$\%$ certainty. The
rotational levels are populated according to the Boltzman-distribution presented in
Fig.~\ref{fig:Selectivity}. We calculate the confidence $F'$ for the molecular ion to be
found in a specific rotational state after application of the state preparation and state
purification scheme. The results presented in Fig.~\ref{fig:Selectivity} show that the
ion can be prepared in all 11 represented rotational states with high confidence. Hence,
if the aim is to study the dependence of a process on the internal state of a molecule,
starting out with an internally hot molecular ion may turn out to be advantageous.

The above presented scheme is not restricted to diatomic molecular ions. For more complex
molecules at thermal equilibrium, the number of populated internal states increases, and
hence the population of the individual states will eventually be too low to prepare a
molecule in a fully specified quantum state. The scheme could, however, be used to
specify one specific quantum number by a suitable choice of the intermediate state in the
Raman process used to correlate the internal state of the molecule with the external
motional mode.

At a more refined level of manipulation and of interest for quantum information
processing, the two-ion system discussed here may be used to create and study
entanglement between atomic and molecular species.

In conclusion, we presented a method to prepare a single trapped
molecular ion in specific states with high confidence. Of great
prospect for state specific investigations, the probabilistic
nature of the preparation process makes it possible to access a
large number of states within a thermal distribution in a relative
simple experimental way.

\acknowledgments We thank K. M{\o}lmer for discussions. The work
is supported by the Danish Natural Science Research Council, the
Danish National Research Foundation through the Quantum Optics
Center QUANTOP and by the Carlsberg Foundation.


\end{document}